\documentclass{article}
\usepackage{graphicx} 
\usepackage{cite} 
\usepackage{amsmath}
\usepackage{authblk}
\usepackage[papersize={8.5in,11in}, margin=1in]{geometry}
\usepackage{setspace}

\title{cuSOAP: a GPU-accelerated Generator of Smooth Overlap of Atomic Positions Descriptor}

\author[1]{Hongyu Yan}
\author[2] {Yuqing Xia}
\author[3]{Yong Wei}
\author[4]{Minghan Chen}
\author[5*]{Hanning Chen}

\affil[1]{School of Engineering and Physical Sciences, Heriot-Watt University, Edinburgh EH14 4AS, UK}
\affil[2]{School of Artificial Intelligence, Chinese University of Hong Kong, Shatin, Hong Kong}
\affil[3]{Department of Computer Science and Cybersecurity, University of North Georgia, Dahlonega, Georgia 30597, USA}
\affil[4]{Department of Computer Science, Wake Forest University, Winston-Salem, North Carolina 27104, USA}
\affil[5]{Texas Advanced Computing Center, The University of Texas at Austin, Austin, Texas 78758, USA}

\affil[*]{Corresponding authors: hanning.chen@austin.utexas.edu}

\date{}

\begin{document}

\maketitle

\doublespacing

\begin{abstract}
The Smooth Overlap of Atomic Positions (SOAP) descriptor is one of the most
widely adopted representations of atomic environments in molecular machine
learning, but the cost of evaluating it and its derivatives remains a principal bottleneck of SOAP-based
interatomic potentials, particularly at the large radial and angular basis
sizes demanded by complex, multi-species condensed-phase environments. We
present \texttt{cuSOAP}, a GPU-accelerated generator of atom-wise SOAP vectors
and their analytic derivatives. Built on \texttt{PyTorch}, \texttt{cuSOAP}
evaluates closed-form expressions or quadratures for the projection coefficients and their
Cartesian gradients for Gaussian-type-orbital and polynomial radial
bases, through fused CUDA and Triton kernels that eliminate the multi-gigabyte
intermediates of a naive tensor formulation. The package is a drop-in
replacement for the CPU-based reference \texttt{DScribe}, reproducing its
constructor signature, feature ordering, and output to within ${\sim}10^{-6}$,
and accepts structures directly as Atomistic Simulation Environment (ASE)
\texttt{Atoms} objects. On an NVIDIA Grace--Blackwell (GB200) node, a single
Blackwell GPU generates the full descriptor-plus-Jacobian workload for a
1000-molecule water cluster up to two orders of magnitude faster than
\texttt{DScribe}, i.e., from $13\times$ to $227\times$ across the entire
$(n_{\max}, l_{\max})$ hyperparameter map, with the speedup growing with the
angular band limit $l_{\max}$. Descriptor-only generation scales as
$t \propto n^{1.34}$, close to linear, up to a million-atom water cluster,
which is processed in 19.5~s on one GPU and, through a shared-memory
multiprocessing driver, in 5.7~s on the four GPUs of the node, demonstrating $\sim$ 90\% parallel efficiency with no sign
of saturation as devices are added. These results bring on-the-fly SOAP
evaluation for large-scale condensed-phase simulation within reach.
\end{abstract}

\section{Introduction}
\label{sec:introduction} 
The rapid maturation of artificial intelligence has reshaped the molecular and materials sciences, where machine-learning (ML) models now routinely
predict potential energy surfaces, forces, and electronic properties at a fraction of the cost of first-principles calculations.\cite{behler2007,bartok2010,deringer2021} The predictive power and transferability of such models hinge less on the regression algorithm
than on the way an atomic structure is encoded: the mathematical descriptor that maps a set of Cartesian coordinates and chemical identities onto a fixed-length feature vector.\cite{bartok2013,himanen2020} A useful descriptor must be invariant to the symmetries of physics (i.e., translation, rotation, reflection, and permutation of like atoms), while remaining a faithful, differentiable representation of the local environment.\cite{bartok2013,de2016} Among the descriptors developed to meet these requirements, the Smooth Overlap of Atomic Positions (SOAP) has become one of the most widely adopted.\cite{bartok2013,de2016} SOAP represents each atomic neighborhood as a smooth Gaussian-smeared density that is expanded in a basis of radial functions and spherical harmonics; contracting this expansion into the rotationally invariant power spectrum yields a descriptor with several attractive properties. Its invariances are built in by construction rather than learned from data.\cite{bartok2013} Replacing the atomic Dirac deltas with Gaussians makes both the descriptor and the derived similarity kernel vary smoothly and differentiably with the atomic coordinates, so that analytic forces are well defined.\cite{bartok2013} The representation converges systematically toward completeness as a small number of adjustable
hyperparameters such as radial basis size $n_{\max}$, angular band limit $l_{\max}$, and cutoff radius $r_{\mathrm{cut}}$ are increased, providing a controllable trade-off between accuracy and cost.\cite{bartok2013,de2016} The associated SOAP kernel induces a proper, positive-definite metric $d(X,X') = \sqrt{2 - 2\,\hat{p}(X)\cdot\hat{p}(X')}$ on the space of environments, giving a well-defined and physically meaningful measure of structural similarity.\cite{de2016} Finally, because SOAP arises from a body-order expansion of the atomic density, it offers a systematic, many-body route to a complete description of the local environment.\cite{bartok2013}

These strengths, however, come at a substantial computational cost, which is the principal obstacle to applying SOAP to large systems. The dominant drawback is the sheer size and expense of the descriptor. The power spectrum comprises elements indexed by every pair of radial functions, each angular channel, and every ordered pair of chemical species, so its dimension scales as $\mathcal{O}(n_{\max}^{2}\,l_{\max}\,S^{2})$ for $S$ species, and the cost of evaluating it grows steeply with the basis size and with the number of neighbors in the cutoff sphere.\cite{bartok2013,de2016,laakso2023} The analytic derivatives required for forces aggravate this further: they must be accumulated with respect to the position of every neighbor, inflating the cost by a factor of roughly $3N_{\mathrm{neigh}}$.\cite{laakso2023} As a consequence, SOAP-based potentials are typically orders of magnitude slower than simple descriptors or classical force fields. A second difficulty is redundancy. The rotationally invariant power spectrum contains many correlated, near-degenerate components, so that the representation is generically overcomplete;\cite{bartok2013} for large systems the resulting memory footprint becomes prohibitive, and practitioners routinely resort to sparsification or dimensionality reduction. For example, the compression of a $252$-dimensional SOAP vector to $50$ components was needed by a principal-component analysis.\cite{laakso2023} A third challenge is that the accuracy of SOAP is sensitive to its hyperparameters: the cutoff radius, Gaussian width, $n_{\max}$, and $l_{\max}$ must be tuned for each system, and a poorly chosen cutoff hard-limits the physics the descriptor can capture, since it fixes the length scale on which structural differences are resolved.\cite{de2016} Taken together, these factors make an efficient SOAP generator highly desirable, particularly for large systems with complex, multi-species chemical environments.

Graphics processing units (GPUs) are exceptionally well matched to this workload. The generation of SOAP descriptors is embarrassingly parallel over atomic centers, over the neighbors within each cutoff sphere, and over the $(n, l, m)$ and species indices of the expansion coefficients---exactly the kind of massively data-parallel, arithmetically dense computation for which the GPU's thousands of lightweight cores and high-bandwidth memory are designed. In contrast to a CPU, which devotes most of its transistors to control logic and caching for a handful of latency-optimized threads, a GPU dedicates its die area to throughput-optimized arithmetic units, so that the many independent spherical-harmonic evaluations, radial-basis contractions, and per-neighbor derivative accumulations that dominate SOAP can be executed concurrently and their memory traffic hidden behind computation. The regular, predictable access pattern of the power-spectrum contraction maps naturally onto coalesced memory accesses and fast on-chip shared memory, while the derivative computation, which is the most expensive stage on a CPU, exposes precisely the additional parallelism that a GPU thrives on. These hardware advantages are matched by a mature software stack: programming models such as CUDA, together with tensor libraries that expose GPU kernels through Python and automatic-differentiation frameworks, allow high-performance descriptor kernels to be written once and integrated directly into modern ML pipelines,\cite{bigi2023} whereas the established SOAP implementations remain CPU-bound.\cite{himanen2020,laakso2023}

With these considerations in mind, we present \texttt{cuSOAP}, a GPU-accelerated software package for the generation of SOAP descriptors and their analytic derivatives for molecular systems. \texttt{cuSOAP} recasts the descriptor and force evaluation as GPU kernels that exploit the parallelism inherent in the SOAP construction, and exposes them through a convenient Python interface suitable for direct use in machine-learning workflows. In the following sections we present the mathematical formulation underlying the implementation (Sec.~\ref{sec:derivations}), describe the numerical and computational design of the GPU kernels (Sec.~\ref{sec:implementation}), benchmark the accuracy and performance of \texttt{cuSOAP} against existing CPU-based generators (Sec.~\ref{sec:results}), and summarize our conclusions (Sec.~\ref{sec:conclusions}).

\section{Mathematical Derivations}
\label{sec:derivations}

\subsection{Gaussian-smeared atomic density and its basis-set projection}
\label{sec:density}

The SOAP construction begins by replacing the atoms surrounding a given
central atom with smooth Gaussian densities.\cite{bartok2013} For each
chemical species $\mu$, the neighbor density seen by the central atom is
\begin{equation}
  \rho^{\mu}(\mathbf{r})
  = \sum_{i=1}^{N^{\mu}_{a}}
    \exp\!\left(-\frac{\left|\mathbf{r}-\mathbf{r}_{i\mu}\right|^{2}}
                      {2\sigma^{2}}\right),
  \label{eq:density}
\end{equation}
where the sum runs over the $N^{\mu}_{a}$ atoms of species $\mu$ located
within the cutoff radius $r_{\mathrm{cut}}$ of the central atom,
$\mathbf{r}_{i\mu}$ is the position of the $i$-th such atom relative to the
central atom, and $\sigma$ is the width of the smearing Gaussians. Depending
on convention, the central atom itself may be included in the density of its
own species; its contribution is made explicit below. Keeping one density per
species preserves the chemical identity of the neighbors and yields the
familiar multi-species (``partial'') power spectrum.\cite{de2016,himanen2020}

Within the cutoff sphere, the density of Eq.~\eqref{eq:density} is expanded
in a basis of radial functions $g_{nl}(r)$ and spherical harmonics
$Y_{lm}(\theta,\phi)$,
\begin{equation}
  \rho^{\mu}(\mathbf{r})
  = \sum_{n=1}^{n_{\max}} \sum_{l=0}^{l_{\max}} \sum_{m=-l}^{+l}
    c^{\mu}_{nlm}\, g_{nl}(r)\, Y_{lm}(\theta,\phi),
  \label{eq:expansion}
\end{equation}
where $n_{\max}$ is the number of radial functions and $l_{\max}$ the
maximum angular momentum. Following the convention of
\texttt{DScribe},\cite{himanen2020,laakso2023} the $Y_{lm}$ are taken to be
the real spherical harmonics, so that all expansion coefficients are real and
no complex conjugation needs to be tracked. If the radial functions are
orthonormal for each angular channel, the expansion coefficients are obtained
by direct projection,
\begin{equation}
  c^{\mu}_{nlm}
  = \int_{0}^{r_{\mathrm{cut}}}\! \mathrm{d}r\, r^{2}
    \int \mathrm{d}\Omega\;
    g_{nl}(r)\, Y_{lm}(\theta,\phi)\,
    \rho^{\mu}(r,\theta,\phi),
  \label{eq:projection}
\end{equation}
where the angular integration extends over the full solid angle,
$\mathrm{d}\Omega=\sin\theta\,\mathrm{d}\theta\,\mathrm{d}\phi$.

The rotationally invariant SOAP descriptor is assembled from the
coefficients of Eq.~\eqref{eq:projection} as the partial power
spectrum\cite{bartok2013,laakso2023}
\begin{equation}
  p^{\mu\nu}_{nn'l}
  = \pi\sqrt{\frac{8}{2l+1}}
    \sum_{m=-l}^{+l} c^{\mu}_{nlm}\, c^{\nu}_{n'lm},
  \label{eq:powerspectrum}
\end{equation}
collected over all species pairs $(\mu,\nu)$ and all unique combinations of
the basis-function indices $(n,n',l)$; this is the origin of the
$\mathcal{O}(n_{\max}^{2}\,l_{\max}\,S^{2})$ scaling of the descriptor
dimension discussed in Sec.~\ref{sec:introduction}. Because
Eq.~\eqref{eq:powerspectrum} is a simple quadratic contraction, the
derivatives of the descriptor with respect to the position
$\mathbf{R}_{j}$ of any atom $j$ follow from the product
rule,\cite{laakso2023}
\begin{align}
  \nabla_{\!j}\, p^{\mu\nu}_{nn'l}
  = \pi\sqrt{\frac{8}{2l+1}}
    \sum_{m=-l}^{+l}
    \Bigl[&
      \left(\nabla_{\!j}\, c^{\mu}_{nlm}\right) c^{\nu}_{n'lm}
  \nonumber\\
    &+ c^{\mu}_{nlm} \left(\nabla_{\!j}\, c^{\nu}_{n'lm}\right)
    \Bigr],
  \label{eq:powerspectrum-derivative}
\end{align}
so that the entire problem of evaluating SOAP vectors and the analytic
derivatives needed for forces reduces to obtaining closed-form expressions
for $c^{\mu}_{nlm}$ and $\nabla c^{\mu}_{nlm}$. Since the coefficients are
functions of the interatomic separations only, translational invariance
implies that the derivative with respect to the central atom is the negative
of the sum over its neighbors' derivatives, and only atoms inside the cutoff
sphere contribute nonzero terms. In the remainder of this section we derive
such closed forms for the two radial-basis families implemented in
\texttt{cuSOAP}: Gaussian-type orbitals (Sec.~\ref{sec:gto}) and cubic and
higher-order polynomials (Sec.~\ref{sec:poly}).

\subsection{Gaussian-type-orbital radial functions}
\label{sec:gto}

The first family of radial functions consists of linear combinations of
spherical Gaussian-type orbitals (GTOs),\cite{himanen2020,laakso2023}
\begin{equation}
  g_{nl}(r) = \sum_{k=1}^{n_{\max}} \beta^{l}_{nk}\, r^{l}
              e^{-\alpha_{kl} r^{2}},
  \label{eq:gto-basis}
\end{equation}
where the $\alpha_{kl}$ are fixed primitive widths and the coefficients
$\beta^{l}_{nk}$ orthonormalize the basis within each angular channel $l$.
The primitive overlap matrix required for the orthonormalization is
available in closed form,
\begin{align}
  S^{l}_{kk'}
  &= \int_{0}^{r_{\mathrm{cut}}} \mathrm{d}r\;
     r^{2l+2}\, e^{-(\alpha_{kl}+\alpha_{k'l})r^{2}}
  \nonumber\\
  &= \frac{\Gamma\!\left(l+\tfrac{3}{2}\right)
           - \Gamma\!\left(l+\tfrac{3}{2},
             (\alpha_{kl}+\alpha_{k'l})\,r_{\mathrm{cut}}^{2}\right)}
          {2\,(\alpha_{kl}+\alpha_{k'l})^{\,l+3/2}},
  \label{eq:gto-overlap}
\end{align}
where $\Gamma(s)$ and $\Gamma(s,x)$ denote the gamma function and the upper
incomplete gamma function, respectively. The widths are chosen so that every
primitive has decayed essentially to zero at the cutoff,
$\alpha_{kl} r_{\mathrm{cut}}^{2} \gg 1$; the incomplete-gamma correction in
Eq.~\eqref{eq:gto-overlap} is then negligible and
\begin{equation}
  S^{l}_{kk'} \approx
  \frac{\Gamma\!\left(l+\tfrac{3}{2}\right)}
       {2\,(\alpha_{kl}+\alpha_{k'l})^{\,l+3/2}}.
  \label{eq:gto-overlap-approx}
\end{equation}
The orthonormalization coefficients follow by symmetric (L\"owdin)
orthogonalization,
\begin{equation}
  \boldsymbol{\beta}^{l} = \left(\mathbf{S}^{l}\right)^{-1/2}.
  \label{eq:gto-lowdin}
\end{equation}
For the primitive widths, \texttt{cuSOAP} adopts the convention of
\texttt{DScribe}~2.1,\cite{laakso2023} in which the $k$-th primitive of
angular momentum $l$ decays to a threshold value of $10^{-3}$ at a distance
$z_{k}$, with the $z_{k}$ spaced uniformly between $1~\text{\AA}$ and the
cutoff radius:
\begin{equation}
  \alpha_{kl} = -\frac{\ln\!\left(10^{-3}/z_{k}^{\,l}\right)}{z_{k}^{2}},
  \qquad
  z_{k} = 1 + (k-1)\,\frac{r_{\mathrm{cut}}-1}{n_{\max}-1},
  \label{eq:gto-alphas}
\end{equation}
with distances expressed in {\aa}ngstr\"om.

The virtue of the GTO basis is that the projection integral of
Eq.~\eqref{eq:projection} can be evaluated in closed form. Substituting
Eqs.~\eqref{eq:density} and \eqref{eq:gto-basis} into
Eq.~\eqref{eq:projection}, each neighbor--primitive pair contributes a
three-dimensional Gaussian integral against $r^{l}Y_{lm}$. This integral is
evaluated by expanding the off-center smearing Gaussian in modified
spherical Bessel functions of the first kind, $i_{l}(z)$,
\begin{align}
  \exp\!\left(-\frac{\left|\mathbf{r}-\mathbf{r}'\right|^{2}}
                    {2\sigma^{2}}\right)
  ={}& 4\pi
     \exp\!\left(-\frac{r^{2}+r'^{2}}{2\sigma^{2}}\right)
  \nonumber\\
  &\times
     \sum_{l=0}^{\infty}\sum_{m=-l}^{+l}
     i_{l}\!\left(\frac{r r'}{\sigma^{2}}\right)
     Y_{lm}(\hat{\mathbf{r}})\, Y_{lm}(\hat{\mathbf{r}}'),
  \label{eq:bessel-expansion}
\end{align}
whereupon the angular integral selects the $(l,m)$ channel and the remaining
radial integral is elementary once the upper limit is extended to infinity
(again permissible because $\alpha_{kl} r_{\mathrm{cut}}^{2}\gg 1$),
\begin{equation}
  \int_{0}^{\infty}\!\mathrm{d}r\; r^{\,l+2}\, e^{-p r^{2}}\,
  i_{l}(c\,r)
  = \frac{\sqrt{\pi}\; c^{\,l}}{2^{\,l+2}\, p^{\,l+3/2}}\,
    e^{c^{2}/4p},
  \label{eq:bessel-integral}
\end{equation}
with $p = \alpha_{kl} + 1/(2\sigma^{2})$ and $c = r_{i\mu}/\sigma^{2}$.
Collecting all prefactors yields the closed-form projection coefficients
\begin{align}
  c^{\mu}_{nlm}
  ={}& (2\pi)^{3/2}\sigma^{3}
     \sum_{i=1}^{N^{\mu}_{a}} \sum_{k=1}^{n_{\max}}
     \beta^{l}_{nk}\,
     \frac{r_{i\mu}^{\,l}\, Y_{lm}(\hat{\mathbf{r}}_{i\mu})}
          {\left(1+2\sigma^{2}\alpha_{kl}\right)^{l+3/2}}
  \nonumber\\
  &\times
     \exp\!\left(-\frac{\alpha_{kl}\, r_{i\mu}^{2}}
                       {1+2\sigma^{2}\alpha_{kl}}\right)
  \nonumber\\
  &+ \delta_{l0}\,\delta_{m0}\,
     \sqrt{2}\,\pi\sigma^{3}
     \sum_{k=1}^{n_{\max}}
     \frac{\beta^{0}_{nk}}
          {\left(1+2\sigma^{2}\alpha_{k0}\right)^{3/2}},
  \label{eq:gto-coefficients}
\end{align}
where $r_{i\mu}=|\mathbf{r}_{i\mu}|$ and
$\hat{\mathbf{r}}_{i\mu}=\mathbf{r}_{i\mu}/r_{i\mu}$. The second term is the
on-site contribution of the central atom itself, obtained from the first
term in the limit $r_{i\mu}\to 0$, where only the $l=0$, $m=0$ channel
survives ($r^{l}Y_{lm}\to\delta_{l0}\delta_{m0}/\sqrt{4\pi}$ as $r\to0$); it
is present only when the central atom is of species $\mu$ and is included in
the neighbor density. Because this term is independent of all atomic
positions, it shifts the $l=0$ coefficients by a constant and does not
contribute to any derivative.

Equation~\eqref{eq:gto-coefficients} is smooth in the neighbor coordinates,
and its Cartesian gradients follow from the product rule applied to the
solid harmonic $r^{l}Y_{lm}(\hat{\mathbf{r}})$ and to the Gaussian envelope.
Using the gradient recursion for solid harmonics, the derivative with
respect to the $x$ coordinate of neighbor $j$ of species $\mu$ reads

\begin{align}
  \frac{\partial c^{\mu}_{nlm}}{\partial x_{j\mu}}
  ={}& (2\pi)^{3/2}\sigma^{3}
    \sum_{k=1}^{n_{\max}}
    \frac{\beta^{l}_{nk}\; r_{j\mu}^{\,l-1}}
         {\left(1+2\sigma^{2}\alpha_{kl}\right)^{l+3/2}}\,
    \exp\!\left(-\frac{\alpha_{kl}\, r_{j\mu}^{2}}
                      {1+2\sigma^{2}\alpha_{kl}}\right)
  \Biggl[
    \frac{1}{2}\sqrt{\frac{(l-m)(l-m-1)(2l+1)}{2l-1}}\;
    Y_{l-1,m+1}(\hat{\mathbf{r}}_{j\mu})
  \nonumber\\
  &\qquad
    - \frac{1}{2}\sqrt{\frac{(l+m)(l+m-1)(2l+1)}{2l-1}}\;
    Y_{l-1,m-1}(\hat{\mathbf{r}}_{j\mu})
    - \frac{2\alpha_{kl}\, x_{j\mu}\, r_{j\mu}}
         {1+2\sigma^{2}\alpha_{kl}}\;
    Y_{lm}(\hat{\mathbf{r}}_{j\mu})
  \Biggr],
  \label{eq:gto-derivative}
\end{align}

with entirely analogous expressions for the $y$ and $z$ components. The
first two terms of the bracket are the gradient of the solid harmonic, here
written in terms of the complex spherical harmonics for compactness; the
corresponding relations for the real harmonics used in the implementation
are obtained by taking the appropriate real and imaginary combinations, or
equivalently the Cartesian derivatives of $Y_{lm}$ can be generated directly
by stable hardware-friendly recursions of the kind implemented in
\texttt{sphericart}.\cite{bigi2023} The last term arises from the Gaussian
envelope. Only the term $i=j$ of Eq.~\eqref{eq:gto-coefficients} survives
differentiation, so the cost of a single coefficient derivative is
independent of the neighbor count, while the full derivative array spans all
$3N_{\mathrm{neigh}}$ neighbor coordinates, as anticipated in
Sec.~\ref{sec:introduction}.

\subsection{Cubic and higher-order polynomial radial functions}
\label{sec:poly}

The second family of radial functions is built from polynomials that vanish
smoothly at the cutoff,\cite{himanen2020}
\begin{equation}
  g_{n}(r) = \sum_{k=1}^{n_{\max}}
             W_{nk}\,\frac{\left(r_{\mathrm{cut}}-r\right)^{k+2}}{N_{k}},
  \label{eq:poly-basis}
\end{equation}
where the lowest member is cubic, guaranteeing that every basis function and
its first two radial derivatives vanish at $r=r_{\mathrm{cut}}$. In contrast
to the GTO basis, the polynomial radial functions are independent of the
angular channel. The normalization factors are
\begin{equation}
  N_{k} = \sqrt{\int_{0}^{r_{\mathrm{cut}}}
          \left(r_{\mathrm{cut}}-r\right)^{2(k+2)} \mathrm{d}r}
        = \sqrt{\frac{r_{\mathrm{cut}}^{\,2k+5}}{2k+5}},
  \label{eq:poly-norm}
\end{equation}
and the overlap matrix of the normalized primitives is
\begin{align}
  S_{kk'} &= \int_{0}^{r_{\mathrm{cut}}} \mathrm{d}r\,
            \frac{\left(r_{\mathrm{cut}}-r\right)^{k+2}
                  \left(r_{\mathrm{cut}}-r\right)^{k'+2}}{N_{k}N_{k'}}
  \nonumber\\
          &= \frac{\sqrt{(5+2k)(5+2k')}}{5+k+k'},
  \label{eq:poly-overlap}
\end{align}
so that the orthonormalization coefficients again follow from L\"owdin
orthogonalization, $\mathbf{W} = \mathbf{S}^{-1/2}$.

Inserting Eqs.~\eqref{eq:density} and \eqref{eq:poly-basis} into
Eq.~\eqref{eq:projection} and using the expansion of
Eq.~\eqref{eq:bessel-expansion}, the angular integrals are again evaluated
analytically, leaving one radial integral per neighbor, primitive, and
angular channel:
\begin{equation}
  c^{\mu}_{nlm}
  = 4\pi \sum_{i=1}^{N^{\mu}_{a}}
    e^{-r_{i\mu}^{2}/2\sigma^{2}}\,
    Y_{lm}(\hat{\mathbf{r}}_{i\mu})
    \sum_{k=1}^{n_{\max}} \frac{W_{nk}}{N_{k}}\,
    I_{kl}(r_{i\mu}),
  \label{eq:poly-coefficients-exact}
\end{equation}
with
\begin{equation}
  I_{kl}(r_{i\mu})
  = \int_{0}^{r_{\mathrm{cut}}} \mathrm{d}r\;
    r^{2}\left(r_{\mathrm{cut}}-r\right)^{k+2}
    e^{-r^{2}/2\sigma^{2}}\,
    i_{l}\!\left(\frac{r\, r_{i\mu}}{\sigma^{2}}\right).
  \label{eq:poly-radial-integral}
\end{equation}
Unlike the GTO case, the integrals $I_{kl}$ possess no elementary closed
form. They are, however, dominated by a sharp Gaussian peak at
$r = r_{i\mu}$: combining the large-argument form of the Bessel function,
$i_{l}(z) \simeq (e^{z}/2z)\bigl[1 - l(l+1)/(2z)\bigr]$, with a Laplace
(steepest-descent) expansion of the remaining integrand about $r_{i\mu}$
gives, to first order in $\sigma^{2}$,
\begin{align}
  I_{kl}(r_{i\mu})
  \approx{}& \sqrt{\frac{\pi}{2}}\,\sigma^{3}\,
     e^{\,r_{i\mu}^{2}/2\sigma^{2}}
     \left(r_{\mathrm{cut}}-r_{i\mu}\right)^{k+2}
  \nonumber\\
  &\times\Biggl[
     1 - \frac{l(l+1)\,\sigma^{2}}{2 r_{i\mu}^{2}}
       - \frac{(k+2)\,\sigma^{2}}
              {r_{i\mu}\left(r_{\mathrm{cut}}-r_{i\mu}\right)}
  \nonumber\\
  &\qquad
       + \frac{(k+2)(k+1)\,\sigma^{2}}
              {2\left(r_{\mathrm{cut}}-r_{i\mu}\right)^{2}}
     \Biggr].
  \label{eq:poly-laplace}
\end{align}
The exponential factor cancels exactly against the prefactor
$e^{-r_{i\mu}^{2}/2\sigma^{2}}$ of
Eq.~\eqref{eq:poly-coefficients-exact}, and with
$4\pi\sqrt{\pi/2} = (2\pi)^{3/2}$ the projection coefficients become
\begin{align}
  c^{\mu}_{nlm}
  \approx{}& (2\pi)^{3/2}\sigma^{3}
    \sum_{i=1}^{N^{\mu}_{a}}
    Y_{lm}(\hat{\mathbf{r}}_{i\mu})
    \sum_{k=1}^{n_{\max}} \frac{W_{nk}}{N_{k}}
    \left(r_{\mathrm{cut}}-r_{i\mu}\right)^{k+2}
  \nonumber\\
  &\times\Biggl[
     1 - \frac{l(l+1)\,\sigma^{2}}{2 r_{i\mu}^{2}}
       - \frac{(k+2)\,\sigma^{2}}
              {r_{i\mu}\left(r_{\mathrm{cut}}-r_{i\mu}\right)}
  \nonumber\\
  &\qquad
       + \frac{(k+2)(k+1)\,\sigma^{2}}
              {2\left(r_{\mathrm{cut}}-r_{i\mu}\right)^{2}}
     \Biggr].
  \label{eq:poly-coefficients}
\end{align}
The expansion is asymptotic in the ratio of $\sigma$ to the distances
$r_{i\mu}$ and $r_{\mathrm{cut}}-r_{i\mu}$, and is therefore accurate in the
physically relevant regime in which the smearing width is small compared
with both the interatomic separations and the distance of each neighbor from
the cutoff sphere. The on-site contribution of the central atom, for which
$r_{i\mu}=0$ and Eq.~\eqref{eq:poly-laplace} does not apply, affects only
the $l=0$ channel [$i_{l}(0)=\delta_{l0}$] and can be evaluated once from
the one-dimensional integral $I_{k0}(0)$; like its GTO counterpart, it is
independent of the atomic positions and does not contribute to the
derivatives.

Differentiating the radial part of Eq.~\eqref{eq:poly-coefficients} with
respect to the neighbor distance, and retaining terms through
$\mathcal{O}(\sigma^{2})$, gives
\begin{align}
  \frac{\partial c^{\mu}_{nlm}}{\partial r_{j\mu}}
  = (2\pi)^{3/2}\sigma^{3}\,
    Y_{lm}(\hat{\mathbf{r}}_{j\mu})
    \sum_{k=1}^{n_{\max}} \frac{W_{nk}}{N_{k}}
    \left(r_{\mathrm{cut}}-r_{j\mu}\right)^{k+1}
  \Biggl[&
    -(k+2)
    + \frac{l(l+1)\,\sigma^{2}
            \left(r_{\mathrm{cut}}-r_{j\mu}\right)}{r_{j\mu}^{3}}
    + \frac{(k+2)\bigl[l(l+1)+2\bigr]\sigma^{2}}{2 r_{j\mu}^{2}}
  \nonumber\\
  &+ \frac{(k+1)(k+2)\,\sigma^{2}}
          {r_{j\mu}\left(r_{\mathrm{cut}}-r_{j\mu}\right)}
   - \frac{k(k+1)(k+2)\,\sigma^{2}}
          {2\left(r_{\mathrm{cut}}-r_{j\mu}\right)^{2}}
  \Biggr].
  \label{eq:poly-radial-derivative}
\end{align}
The full Cartesian gradient then follows from the chain rule. Writing the
contribution of neighbor $j$ to Eq.~\eqref{eq:poly-coefficients} as
$Y_{lm}(\hat{\mathbf{r}}_{j\mu})\, f_{nl}(r_{j\mu})$, with $f_{nl}$ the
radial factor whose derivative is given by
Eq.~\eqref{eq:poly-radial-derivative},
\begin{equation}
  \frac{\partial c^{\mu}_{nlm}}{\partial x_{j\mu}}
  = \frac{x_{j\mu}}{r_{j\mu}}\,
    Y_{lm}(\hat{\mathbf{r}}_{j\mu})\,
    \frac{\partial f_{nl}}{\partial r_{j\mu}}
  + f_{nl}(r_{j\mu})\,
    \frac{\partial Y_{lm}(\hat{\mathbf{r}}_{j\mu})}{\partial x_{j\mu}},
  \label{eq:poly-cartesian-derivative}
\end{equation}
and analogously for $y_{j\mu}$ and $z_{j\mu}$, where the Cartesian
derivatives of the spherical harmonics are evaluated with the same
recursions used in the GTO case [cf.\
Eq.~\eqref{eq:gto-derivative}].\cite{bigi2023} Equations
\eqref{eq:gto-coefficients}--\eqref{eq:gto-derivative} and
\eqref{eq:poly-coefficients}--\eqref{eq:poly-cartesian-derivative}, combined
with the contractions of Eqs.~\eqref{eq:powerspectrum} and
\eqref{eq:powerspectrum-derivative}, constitute the complete closed-form
working equations implemented in \texttt{cuSOAP}. Their evaluation is
independent for every (center, neighbor, $n$, $l$, $m$, species) tuple, a
structure that maps directly onto the massively parallel execution model of
the GPU; the corresponding kernel design is described in
Sec.~\ref{sec:implementation}.

\section{Numerical Implementations}
\label{sec:implementation}

The closed-form working equations of Sec.~\ref{sec:derivations} are exact but,
by themselves, say nothing about how the descriptor should be laid out in
memory or scheduled across the arithmetic units of a GPU. This section
describes how \texttt{cuSOAP} turns those equations into a fast,
GPU-resident implementation. We first outline the software architecture and
the user-facing interface (Sec.~\ref{sec:architecture}), then describe the
custom CUDA and Triton kernels that carry the cost of the derivative
evaluation (Sec.~\ref{sec:cuda}), and finally the shared-memory
multiprocessing layer that distributes the work across several GPUs
(Sec.~\ref{sec:multigpu}).

\subsection{Architecture and interface}
\label{sec:architecture}

\texttt{cuSOAP} is built on \texttt{PyTorch}, which provides the
tensor abstraction, the device-agnostic dispatch onto CUDA, and the
reverse-mode automatic-differentiation engine on which one of the two
derivative backends rests. The entire descriptor pipeline---neighbor search,
spherical-harmonic evaluation, radial-basis contraction, coefficient
accumulation, and the power-spectrum assembly---executes on the GPU, so that a
structure supplied as host arrays is transferred to the device once and the
final descriptor or Jacobian is returned without further host--device traffic.

The package is designed as a drop-in replacement for the widely used
\texttt{DScribe} SOAP generator.\cite{himanen2020,laakso2023} The central
\texttt{SOAP} class reproduces the \texttt{DScribe} constructor signature---the
cutoff radius \texttt{r\_cut}, the basis sizes \texttt{n\_max} and
\texttt{l\_max}, the smearing width \texttt{sigma}, the radial-basis selector
\texttt{rbf} (\texttt{"gto"} or \texttt{"polynomial"}), an optional radial
\texttt{weighting}, the \texttt{crossover} flag controlling whether
off-diagonal species pairs are retained, the \texttt{average} mode, the
\texttt{species} list, and the \texttt{periodic} and \texttt{sparse}
flags---augmented only by the arguments needed to steer the hardware backend
(\texttt{device}, the internal quadrature order \texttt{quad\_n}, and an
optional cap \texttt{max\_num\_neighbors} on the neighbor count). Descriptors
are produced by \texttt{create()} (aliased to \texttt{generate()}), which
returns one feature row per requested center as an
$(n_{\mathrm{centers}}, n_{\mathrm{features}})$ tensor, and derivatives by
\texttt{derivatives()}, which returns an
$(n_{\mathrm{centers}}, n_{\mathrm{atoms}}, 3, n_{\mathrm{features}})$ tensor
in the \texttt{DScribe} index convention. Both accept an explicit list of
centers and an \texttt{include}/\texttt{exclude} atom selection, and the
feature ordering---the nested loop over species pairs $(\mu,\nu)$, the
radial-index triangle $(n,n')$, and the angular channel $l$---is chosen to
match \texttt{DScribe} element for element, so that existing analysis and
machine-learning pipelines can adopt \texttt{cuSOAP} without changing the
downstream code. The four averaging modes of \texttt{DScribe} are all
supported: the per-center power spectrum (\texttt{"off"}), the two
density-averaged variants (\texttt{"inner"} and \texttt{"outer"}), and a
\texttt{"cc"} mode that exposes the raw projection coefficients $c^{\mu}_{nlm}$
themselves rather than their quadratic contraction.

Three numerical choices are central to reproducing \texttt{DScribe}'s output to
near machine precision while running on the GPU. First, although the
user-facing \texttt{dtype} defaults to single precision, all internal
arithmetic is carried out in \texttt{float64} and only the final descriptor or
Jacobian is cast down. This is not gratuitous: the L\"owdin-orthonormalized GTO
basis of Eq.~\eqref{eq:gto-lowdin} is severely ill-conditioned at large
$n_{\max}$, with orthonormalization coefficients that reach $\sim\!10^{6}$ and
whose evaluation involves heavy cancellation, so a single-precision forward
pass would lose roughly four significant digits. Double-precision accumulation
followed by a final cast costs essentially nothing at these problem sizes and
brings the descriptor to within $\sim\!10^{-6}$ of the reference. Second, the
angular functions are evaluated with \texttt{sphericart},\cite{bigi2023} which
supplies the real spherical harmonics used in the forward pass and, for the
analytic derivatives, the real \emph{solid} harmonics $r^{l}Y_{lm}$ together
with their exact Cartesian gradients through the stable, hardware-friendly
recursions anticipated in Sec.~\ref{sec:derivations}. Because
\texttt{sphericart} guarantees
$r^{l}Y_{lm}(\hat{\mathbf{r}}) = |\mathbf{r}|^{l}\,Y_{lm}(\mathbf{r}/|\mathbf{r}|)$
by construction, the harmonics entering the coefficients and those entering
their gradients are mutually consistent to machine precision. Third, the
basis-set constants are precomputed once at construction time: the GTO
primitive widths and L\"owdin coefficients of
Eqs.~\eqref{eq:gto-alphas} and \eqref{eq:gto-lowdin}, and, for the polynomial
basis, the overlap matrix and its inverse square root---the latter evaluated,
where \texttt{SciPy} is available, with exactly the same
\texttt{sqrtm}$\circ$\texttt{inv} routine as \texttt{DScribe} so that the
ill-conditioned $\mathbf{S}^{-1/2}$ agrees direction for direction with the
reference. Radial integrals that lack a closed form in the polynomial case are
handled by a fixed Gauss--Legendre quadrature on $[0, r_{\mathrm{cut}}]$.

Neighbor lists are built on the device. When \texttt{torch\_cluster} is
installed, its GPU radius search is used directly; otherwise \texttt{cuSOAP}
falls back to a chunked distance evaluation that rewrites the squared
separation as
$\lVert\mathbf{c}\rVert^{2} + \lVert\mathbf{n}\rVert^{2} - 2\,\mathbf{c}\cdot\mathbf{n}$
so that the dominant cost becomes a single matrix--matrix product handled by
\texttt{cuBLAS}, with the intermediate distance buffer reused across chunks to
limit allocation. For periodic systems the cell and periodic-boundary flags are
honored by extending the neighbor set with the minimal set of periodic images
within the padded cutoff. Because \texttt{sphericart} leaves the angular part
undefined at the origin, edges of vanishing length---which arise when a center
coincides with an atom---are dropped from the neighbor list and their exact
analytic contribution is added separately: this is the on-site $l=0$ self term
of Eqs.~\eqref{eq:gto-coefficients} and \eqref{eq:poly-coefficients}, which is
independent of the atomic positions and therefore leaves the forces unchanged.

Finally, on CUDA devices the constructor performs a warm-up pass on a small
dummy system. This triggers the just-in-time compilation of the custom kernels,
the allocation of the \texttt{cuBLAS} handles and \texttt{sphericart} buffers,
and the pre-population of the caching memory allocator with a block large
enough to serve the Jacobian and its workspaces, so that the first genuine
\texttt{create()} or \texttt{derivatives()} call already runs at steady-state
speed rather than paying one-off compilation and allocation costs. A companion
profiler records per-segment timings when detailed performance breakdowns are
required.

\subsection{Acceleration with CUDA and Triton kernels}
\label{sec:cuda}

The forward evaluation of the descriptor maps naturally onto the batched tensor
primitives of \texttt{PyTorch}. Following the coefficient expressions of
Sec.~\ref{sec:derivations}, the radial factors, spherical harmonics, and
smearing envelopes are formed for every neighbor edge at once, contracted over
the primitive index with a single batched matrix multiplication per angular
channel, and accumulated onto their centers with a scatter-add
(\texttt{index\_add\_}). The power spectrum of Eq.~\eqref{eq:powerspectrum}
follows from one contraction over $m$ per angular channel, after which the
rotationally invariant elements are copied into the \texttt{DScribe} feature
layout using a precomputed slice table. Every one of these operations is
data-parallel over the (center, neighbor, $n$, $l$, $m$, species) index tuple,
exactly the structure anticipated at the end of Sec.~\ref{sec:derivations}.

The derivatives are the expensive part, and \texttt{cuSOAP} offers two backends
for them. The general backend makes the entire forward pass differentiable and
obtains the Jacobian by reverse-mode automatic differentiation; it is exact and
covers every combination of radial basis, averaging mode, weighting, and
periodicity, and is used whenever the fast path does not apply. The default
backend is a closed-form evaluation of the analytic derivative equations
themselves. Reverse-mode differentiation of a descriptor with
$n_{\mathrm{features}}$ outputs per center requires on the order of
$n_{\mathrm{centers}}\!\times\!n_{\mathrm{features}}$ vector--Jacobian passes
and the materialization of a large identity seed, whereas the closed-form path
of Eqs.~\eqref{eq:gto-coefficients}--\eqref{eq:gto-derivative} and
\eqref{eq:poly-coefficients}--\eqref{eq:poly-cartesian-derivative} evaluates the
gradient in a single sweep over the neighbor edges. This replaces a
reverse-mode cost that scales with the descriptor dimension by a handful of
contractions, reducing both the run time and the memory footprint by orders of
magnitude for the system sizes of interest (quantified in
Sec.~\ref{sec:results}).

The decisive step is to fuse the assembly of the derivative array into a small
number of custom kernels written in \texttt{Triton} and compiled to PTX,
rather than expressing it as a chain of library tensor operations. The naive
tensor-operation route would materialize a per-edge gradient tensor of shape
$(E, 3, n_{\max}, (l_{\max}+1)^{2})$, scatter it into a still larger
double-precision buffer indexed by center, included atom, and species, and
evaluate the power-spectrum product rule of
Eq.~\eqref{eq:powerspectrum-derivative} through a sequence of per-$l$
contractions with sizable intermediates. For a few-hundred-atom system at
$n_{\max}=10$, $l_{\max}=5$ these intermediates run to several gigabytes and
dominate the cost. The fused kernels avoid them entirely. A key simplification
is exploited: for a single-image, non-periodic radius search each
(center, atom) pair maps to \emph{at most one} neighbor edge, so every output
element is written by exactly one program instance. This removes the need for
atomic accumulation, for a separate scatter buffer, and even for a
zero-initialization pass---each program stores its result once, writing an
explicit zero where the pair lies outside the cutoff or in the wrong species
block.

Three kernels implement this scheme. For the coefficient (\texttt{"cc"})
Jacobian a single kernel computes, per (center, atom, Cartesian direction)
slot, the edge gradient
$\mathrm{d}c/\mathrm{d}x = \mathbf{r}\,(\hat{B}_{nl}Y_{lm}) + B_{nl}\,\nabla Y_{lm}$,
where $B_{nl}$ is the radial factor and $\hat{B}_{nl}$ its Gaussian-envelope
partner [the two bracketed terms of Eq.~\eqref{eq:gto-derivative}], applies the
\texttt{DScribe} coefficient convention on the fly, and stores the result
directly into the final single-precision layout. For the power-spectrum modes
the assembly is split in two. A first kernel precontracts, for every edge, the
solid harmonics and their gradients against that center's coefficients to form
the two intermediates $U = \sum_{m} Y_{lm}\,c$ and
$V = \sum_{m} \nabla Y_{lm}\,c$; a second kernel then evaluates the product
rule of Eq.~\eqref{eq:powerspectrum-derivative} for each output feature as two
fused multiply--adds against $U$ and $V$, so that the inner sum over $m$
disappears from the hot loop altogether---each derivative term costs two
fused multiply--adds rather than a $(2l+1)$-element dot product. The kernel is
launched on a pair-major one-dimensional grid so that the feature blocks
belonging to one (center, atom) pair are launch-adjacent and reuse that pair's
radial and precontracted rows from cache instead of re-reading them from global
memory. All arithmetic inside the kernels is carried out in double precision
and the result is written once as single precision, so that the several-gigabyte
intermediates of the tensor-operation route collapse to little more than the
output tensor itself. When \texttt{Triton} is unavailable, or the request does
not meet the fast-path conditions (CUDA device and single-precision output), a
pure-\texttt{PyTorch} implementation of the same equations serves as a
fallback, and a central-finite-difference backend is retained for validation.

Two portability details are worth noting. The custom kernels depend on the CUDA
assembler \texttt{ptxas}; because the \texttt{Triton}-bundled assembler may not
recognize very recent GPU targets, \texttt{cuSOAP} prefers a system CUDA
toolkit \texttt{ptxas} when one is present. And because the angular functions
carry a large share of the arithmetic, \texttt{sphericart} is expected to be
built with CUDA support so that the harmonics and their gradients are evaluated
on the device rather than on the host.

\subsection{Multi-GPU parallelization by shared-memory multiprocessing}
\label{sec:multigpu}

The generation of SOAP descriptors is embarrassingly parallel across atomic
centers and across independent structures, and \texttt{cuSOAP} exploits this at
the coarse grain to scale beyond a single accelerator. Because each center's
descriptor and its derivative block depend only on the atoms within that
center's cutoff sphere, distinct centers can be assigned to distinct GPUs with
no inter-device communication and no reduction stage.

The multi-GPU driver is built on \texttt{PyTorch}'s shared-memory
multiprocessing. One worker process is launched per visible device using the
\texttt{spawn} start method required for CUDA, and each worker binds itself to
its own \texttt{cuda:}$\langle$\texttt{rank}$\rangle$ device and constructs an
independent \texttt{SOAP} instance there. The work is first enumerated as a flat
list of tasks, each task being one batch of atomic centers drawn from one
structure; the batch size is a tunable that trades kernel-launch overhead
against per-call memory. Tasks are distributed over the workers round-robin by
rank, so that with $W$ workers each device processes every $W$-th batch. This
static, stride-based assignment needs no central scheduler and keeps all
devices busy whenever the number of batches comfortably exceeds the number of
GPUs.

Coordination uses only lightweight shared-memory primitives. Each worker
signals readiness through a shared result queue once its \texttt{SOAP} object is
constructed and its CUDA kernels are warmed up, and then blocks on a shared
start event; the driver releases every worker simultaneously once all have
reported ready. This barrier ensures that the one-off construction and
compilation costs fall outside the timed region, so that the measured wall time
reflects steady-state descriptor generation rather than start-up. Each worker
transfers its finished descriptors and derivatives to host memory and pushes
them onto the result queue, keyed by structure and batch index, and the driver
reassembles the complete output as the results arrive. Because the workload
partitions cleanly and no gradients or coefficients ever cross device
boundaries, the scheme is data-parallel in the strict sense and its scaling is
limited only by load imbalance across the task list and by the host-side cost
of collecting results---both quantified in Sec.~\ref{sec:results}. The same
center-batching mechanism also serves, in the single-GPU case, to bound the
peak memory of a single call, so that arbitrarily large structures can be
processed by streaming their centers through the device in batches.

\section{Results}
\label{sec:results}

All benchmarks reported in this section were executed on a single NVIDIA
Grace--Blackwell (GB200) compute node equipped with two Grace CPUs, four
Blackwell GPUs, and 1.8~TB of memory. The test systems are water clusters
$(\mathrm{H_{2}O})_{n}$ of increasing size, a convenient two-species
($S=2$) benchmark whose neighbor count per cutoff sphere is representative of
condensed-phase simulations. Unless stated otherwise, the descriptors were
generated with the GTO radial basis, a cutoff radius
$r_{\mathrm{cut}} = 10$~\AA, the "average=off" mode, and
$n_{\max} = 7$, $l_{\max} = 3$. This choice of basis size is not arbitrary:
it mimics the largest atomic orbitals of the naturally occurring elements,
whose occupied shells extend to principal quantum number $n = 7$ and to
$f$~orbitals ($l = 3$), so that the benchmark reflects the resolution required
to describe an arbitrary chemical environment rather than a
water-specific minimum. All timings measure the steady-state generation of
descriptors (and, where indicated, of their analytic derivatives) for every
atom in the system: as described in Sec.~\ref{sec:architecture}, the one-off
kernel compilation, buffer allocation, and warm-up costs are incurred at
construction time and excluded from the timed region. The reference CPU
implementation is \texttt{DScribe},\cite{himanen2020,laakso2023} run on the
Grace CPUs of the same node; as noted in Sec.~\ref{sec:architecture}, the
\texttt{cuSOAP} descriptors and Jacobians agree with the \texttt{DScribe}
reference to within $\sim\!10^{-6}$, so the comparison below is between
numerically equivalent outputs.

\subsection{Comparison with DScribe: descriptors and derivatives}
\label{sec:results-dscribe}

\begin{figure}
  \includegraphics[width=\linewidth]{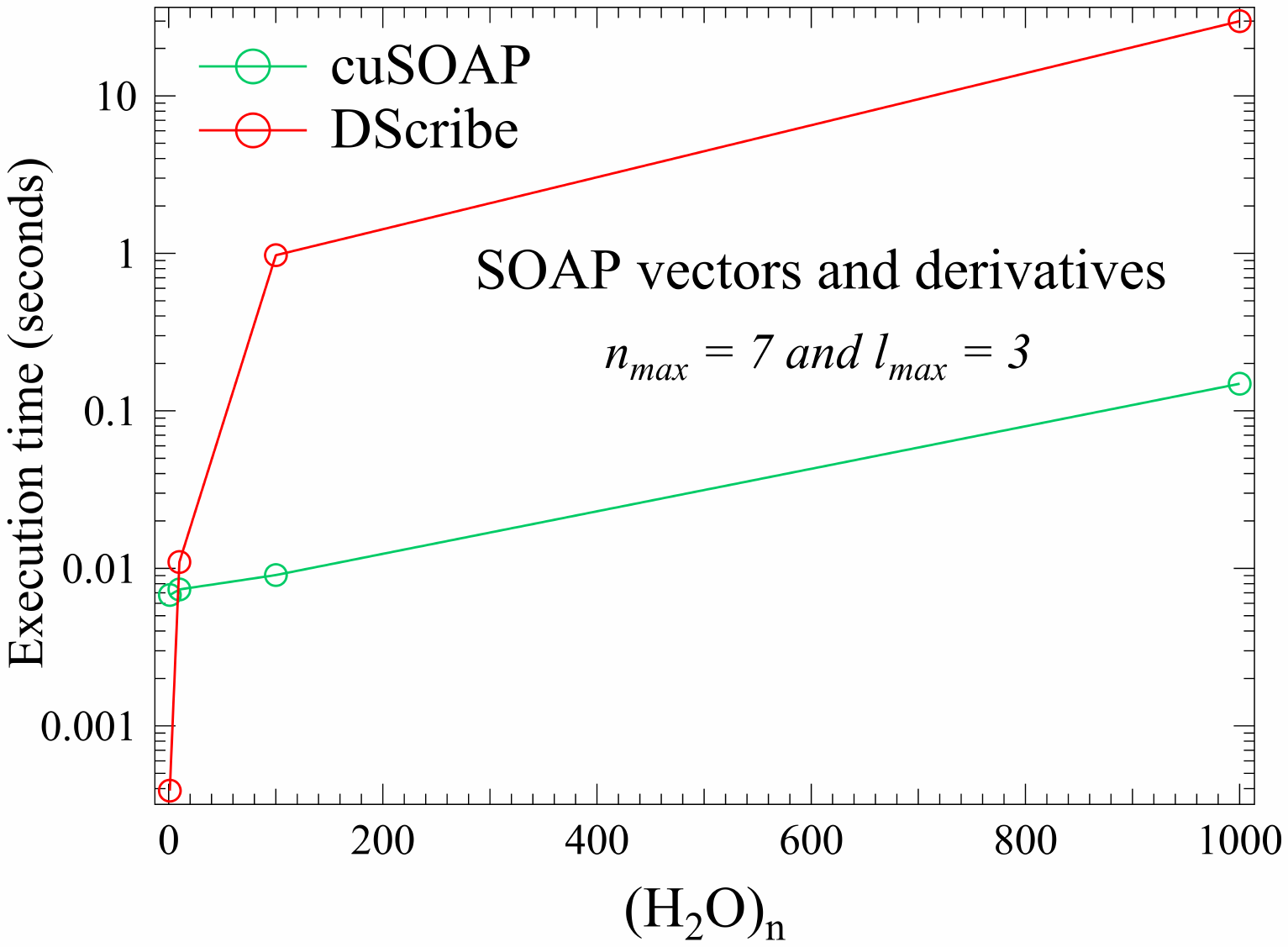}
  \caption{Execution time (seconds, logarithmic scale) for the generation of
  the atom-wise SOAP vectors \emph{and} their analytic derivatives for water
  clusters $(\mathrm{H_{2}O})_{n}$ with $n = 1$--$1000$, using
  $n_{\max} = 7$ and $l_{\max} = 3$. Green: \texttt{cuSOAP} on a single
  Blackwell GPU; red: \texttt{DScribe} on the Grace CPU of the same GB200
  node.}
  \label{fig:dscribe-full}
\end{figure}

Figure~\ref{fig:dscribe-full} compares the time required by \texttt{cuSOAP}
and \texttt{DScribe} to produce the complete output needed for training and
running a machine-learned force field---the per-atom SOAP vectors together
with their derivatives with respect to every atomic coordinate---for clusters
ranging from a single water molecule to $(\mathrm{H_{2}O})_{1000}$. The two
codes show qualitatively different behavior. For the single molecule,
\texttt{DScribe} is in fact faster ($0.4$~ms versus $6.8$~ms): at this size
the GPU computation is entirely dominated by fixed kernel-launch and
synchronization latency, which no amount of parallelism can hide. This
overhead, however, is essentially all there is to the \texttt{cuSOAP} cost
over a wide range of sizes---the GPU time is nearly flat at $7$--$9$~ms from
$n = 1$ to $n = 100$, because the device is latency-bound rather than
throughput-bound until tens of thousands of (center, neighbor) pairs are in
flight---whereas the CPU time grows steeply from the outset. The crossover
occurs already below $n = 10$, where \texttt{cuSOAP} takes the lead
($7.4$~ms versus $11.0$~ms). At $n = 100$ the gap has widened to two orders
of magnitude ($9.1$~ms versus $0.97$~s), and at
$n = 1000$---a cluster of 3000 atoms with roughly $4\times10^{2}$ neighbors
per $10$-\AA{} cutoff sphere---\texttt{cuSOAP} delivers the full descriptor and
Jacobian in $0.15$~s while \texttt{DScribe} requires $29.8$~s: a speedup
factor of $\approx\!200$. The practical consequence is that descriptor
plus force evaluation for a thousand-molecule cluster moves from half a
minute per configuration to a fraction of a second, approaching the cadence
of a molecular-dynamics time step. The
benchmark was capped at $n = 1000$ because the dense Jacobian---one
$3\times n_{\mathrm{features}}$ block per (center, atom) pair---grows
quadratically with system size and, in the \texttt{DScribe} workflow,
exhausts the $1.8$~TB of node memory beyond this point; the streaming
center-batch mechanism of Sec.~\ref{sec:multigpu} lets \texttt{cuSOAP} itself
proceed to larger systems by bounding the per-call output size.

\subsection{Speedup across the hyperparameter space}
\label{sec:results-speedup}

\begin{figure}
  \includegraphics[width=\linewidth]{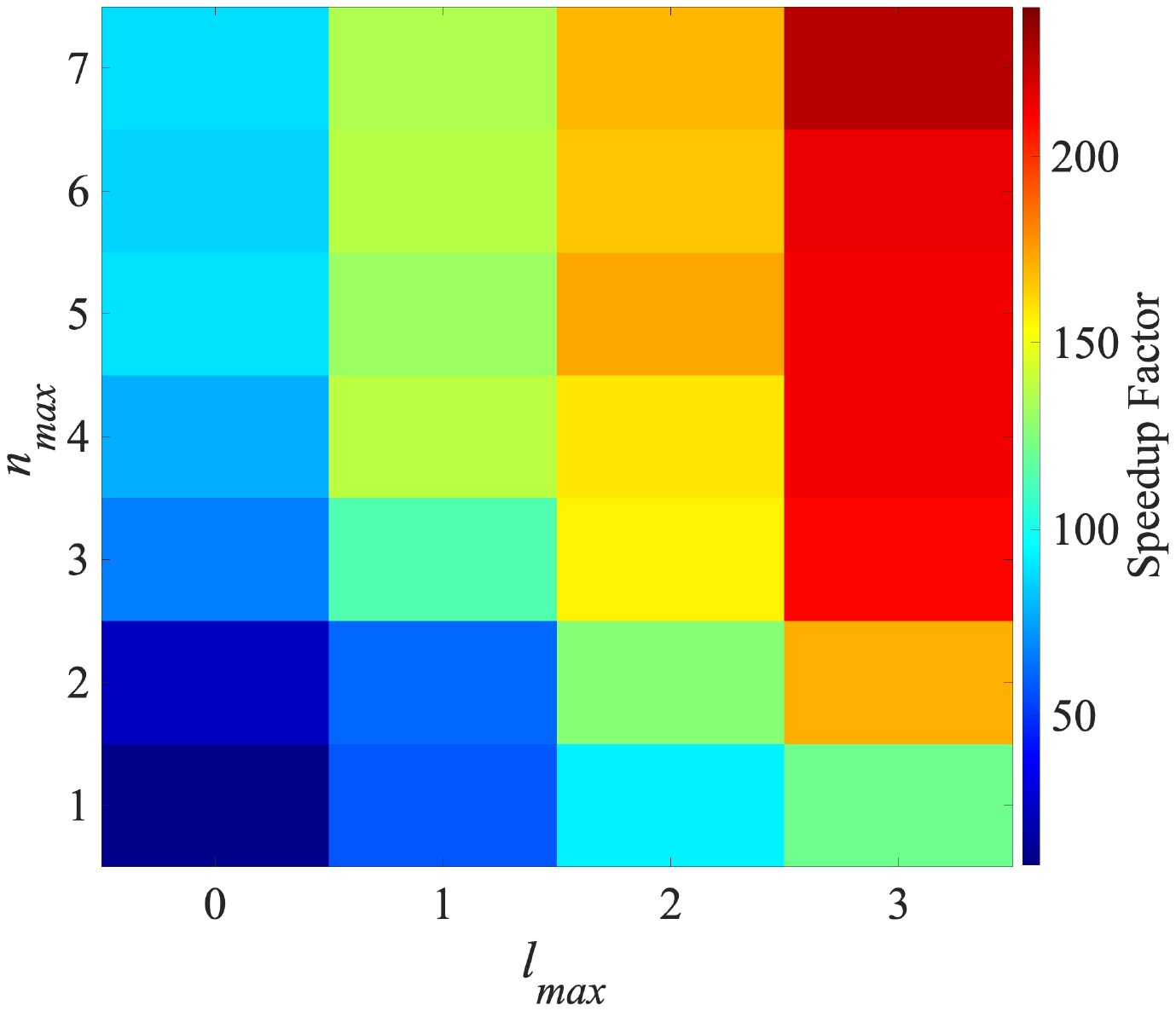}
  \caption{Speedup factor of \texttt{cuSOAP} over \texttt{DScribe} for the
  generation of the atom-wise SOAP vectors and their derivatives for
  $(\mathrm{H_{2}O})_{1000}$, as a function of the radial basis size
  $n_{\max} = 1$--$7$ and the angular band limit $l_{\max} = 0$--$3$.}
  \label{fig:speedup}
\end{figure}

Because the relative cost of the radial contraction, the angular recursion,
and the power-spectrum assembly shifts with the basis-set hyperparameters, a
single $(n_{\max}, l_{\max})$ point does not characterize the acceleration.
Figure~\ref{fig:speedup} therefore maps the speedup factor of \texttt{cuSOAP}
over \texttt{DScribe} for the full descriptor-plus-derivative workload on
$(\mathrm{H_{2}O})_{1000}$ across all 28 combinations of
$n_{\max} = 1$--$7$ and $l_{\max} = 0$--$3$. The speedup grows monotonically
toward the large-basis corner of the map, from $13\times$ in the least
favorable corner, $(n_{\max}, l_{\max}) = (1, 0)$, to $227\times$ at the
largest basis, $(n_{\max}, l_{\max}) = (7, 3)$. Two trends are apparent.
First, at fixed $l_{\max}$ the speedup rises steeply with the radial basis
size up to $n_{\max} \approx 3$ (at $l_{\max} = 3$, from $121\times$ at
$n_{\max} = 1$ to $210\times$ at $n_{\max} = 3$) and then saturates: the
additional radial channels supply the arithmetic intensity needed to fill
the GPU lanes and to amortize the fixed launch overhead, whereas on the CPU
the same channels are paid for at full price from the outset. Second, and
more importantly for applications, at fixed $n_{\max}$ the speedup
\emph{grows} with the angular band limit throughout the map: at
$n_{\max} = 7$ it climbs from $89\times$ at $l_{\max} = 0$ to $135\times$,
$170\times$, and $227\times$ at $l_{\max} = 1$, $2$, and $3$,
respectively. This is precisely the regime in which the
fused-kernel design of Sec.~\ref{sec:cuda} pays off: the
$(l_{\max}+1)^{2}$ angular channels are evaluated by the hardware-friendly
\texttt{sphericart} recursions on the device,\cite{bigi2023} the
precontraction over $m$ removes the $(2l+1)$-element inner products from the
derivative hot loop, and the added angular work largely fills GPU lanes that
would otherwise sit idle, whereas on the CPU the same work is paid for at
full price. High angular resolution is exactly what complex, anisotropic
condensed-phase environments demand of a SOAP model, so the acceleration is
largest where it is needed most: a large $l_{\max}$, prohibitive in a
CPU-bound workflow, becomes essentially free on the GPU.

\subsection{Scaling with system size}
\label{sec:results-scaling}

\begin{figure}
  \includegraphics[width=\linewidth]{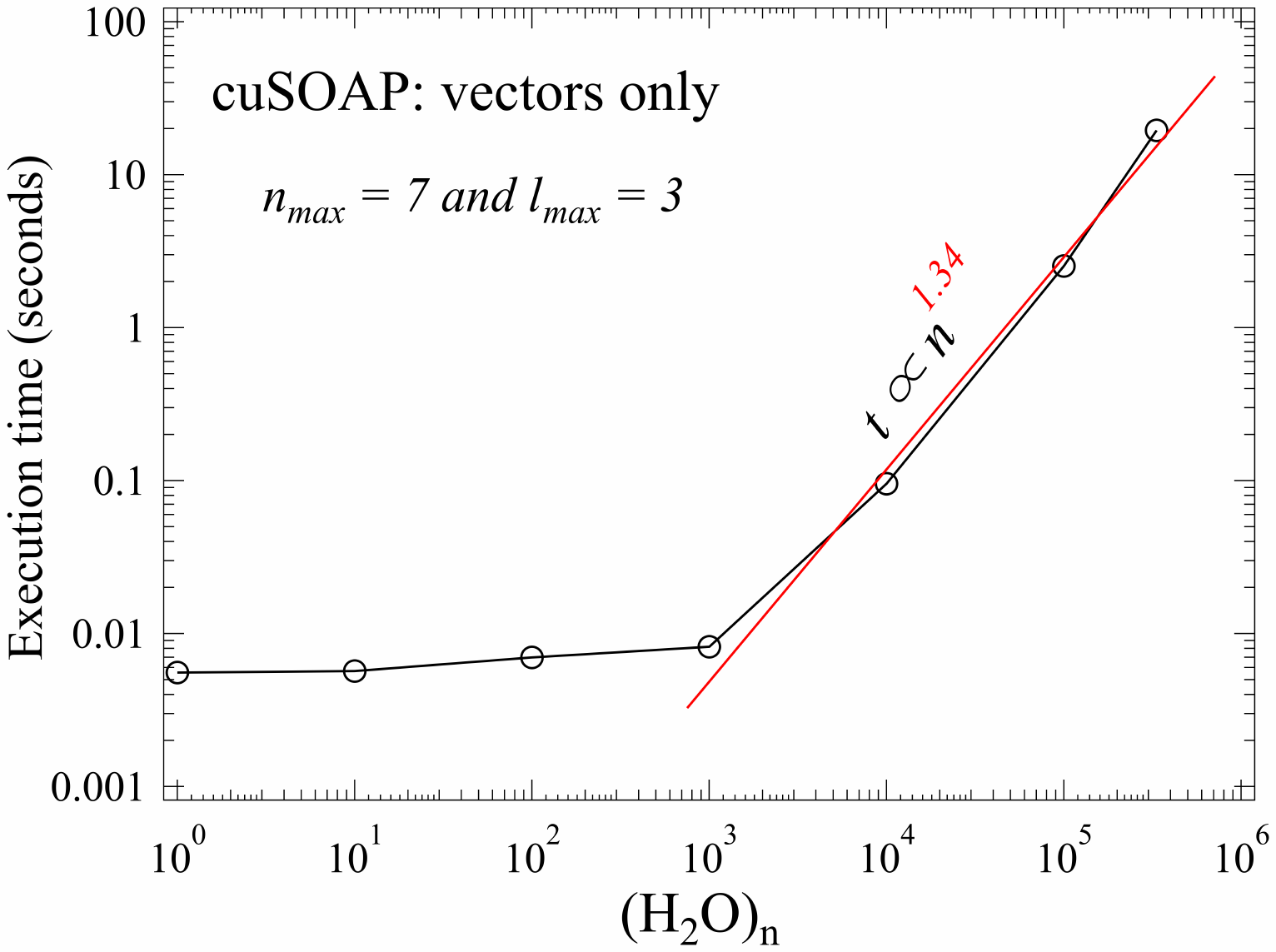}
  \caption{Execution time (seconds) for the generation of the atom-wise SOAP
  vectors (without derivatives) by \texttt{cuSOAP} on a single Blackwell GPU
  for water clusters $(\mathrm{H_{2}O})_{n}$ with $n = 1$--$333\,334$, on a
  double-logarithmic scale. The red line is a power-law fit to the
  large-system regime ($n \geq 10^{3}$), $t \propto n^{1.34}$.}
  \label{fig:vectors-only}
\end{figure}

Many workflows---screening, clustering, kernel-based similarity analysis, and
inference with descriptor-based models that do not require forces---need only
the SOAP vectors themselves. This workload is far lighter than the Jacobian
evaluation in both time and memory: for $(\mathrm{H_{2}O})_{1000}$ the
vectors alone take $8.2$~ms against $0.15$~s for vectors plus derivatives, and
the output grows only linearly, rather than quadratically, with the atom
count. Figure~\ref{fig:vectors-only} probes how far a single GPU can be
pushed in this mode, extending the cluster size from a single molecule to
$(\mathrm{H_{2}O})_{333\,334}$---one million atoms. The timing exhibits the
two regimes anticipated above. Up to $n \approx 10^{3}$ the execution time is
essentially constant at $6$--$8$~ms, pinned by kernel-launch latency: within
this window, systems three orders of magnitude apart in size cost the same,
so the device is effectively free until it saturates. Beyond
$n = 10^{3}$ the time follows a clean power law, $t \propto n^{1.34}$, over
more than two further decades of system size. The exponent is close to the
ideal linear scaling expected for a fixed-density system, in which the number
of neighbors per cutoff sphere is constant and the descriptor work grows
proportionally to the number of centers; the mild super-linearity is
attributable to the pairwise distance screen of the fallback neighbor search
(Sec.~\ref{sec:architecture}), whose cost grows quadratically with the atom
count within each chunk, and it leaves the absolute cost remarkably small: the
million-atom cluster is processed in $19.5$~s on one GPU, and the
$3\times10^{5}$-atom cluster in $2.5$~s.

\subsection{Multi-GPU parallelization for a million-atom system}
\label{sec:results-multigpu}

\begin{figure}
  \includegraphics[width=\linewidth]{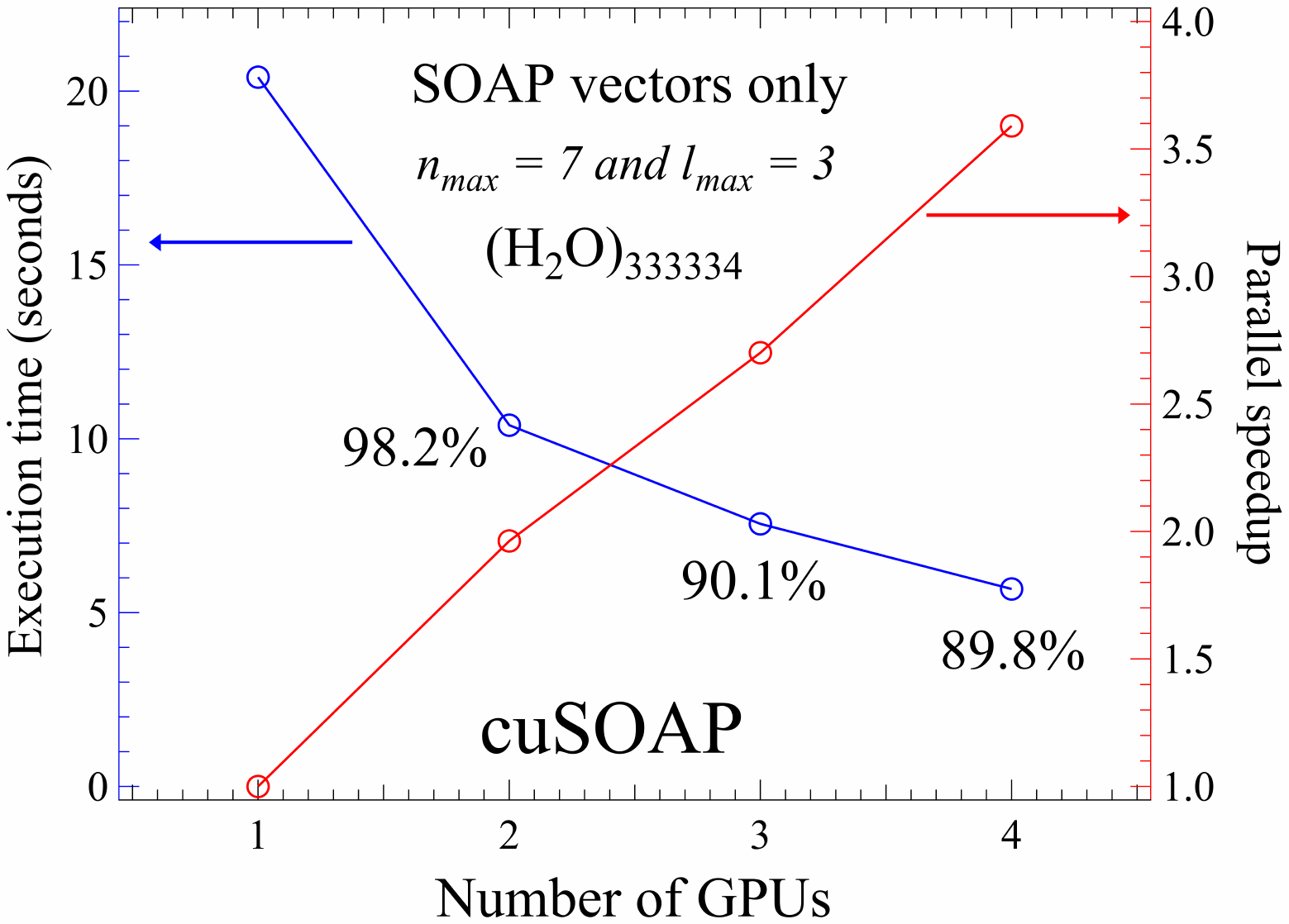}
  \caption{Execution time (seconds; blue, left axis) and parallel speedup
  relative to the single-GPU run (red, right axis) for the generation of the
  atom-wise SOAP vectors for the million-atom cluster
  $(\mathrm{H_{2}O})_{333\,334}$ as a function of the number of Blackwell
  GPUs employed by the shared-memory multiprocessing driver of
  Sec.~\ref{sec:multigpu}. Percentages give the parallel efficiency, i.e.,
  the speedup divided by the number of devices.}
  \label{fig:gpu-scaling}
\end{figure}

Finally, Fig.~\ref{fig:gpu-scaling} examines how the single-node,
shared-memory multiprocessing layer of Sec.~\ref{sec:multigpu} distributes
the million-atom workload of Fig.~\ref{fig:vectors-only} across the four
Blackwell GPUs of the GB200 node. Batches of atomic centers are assigned
round-robin to one worker process per device, with the synchronized-start
barrier ensuring that construction and warm-up fall outside the timed region.
The execution time decreases from $20.4$~s on one GPU to $10.4$~s on two,
$7.6$~s on three, and $5.7$~s on four, and the corresponding parallel
speedup, shown on the right axis of Fig.~\ref{fig:gpu-scaling}, climbs
monotonically to $1.96\times$, $2.70\times$, and $3.59\times$---parallel
efficiencies of $98.2\%$, $90.1\%$, and $89.8\%$, respectively. The
near-ideal two-GPU speedup confirms that the center-partitioned workload
requires no inter-device communication, exactly as the data-parallel
analysis of Sec.~\ref{sec:multigpu} predicts; the modest departure from the
ideal slope at three and four devices is consistent with the residual load
imbalance of the static round-robin assignment and with the serialized
host-side collection of results through the shared queue, both of which grow
in relative weight as the per-device compute time shrinks. Notably, the
speedup curve shows no sign of saturation up to the four devices available
on the node: each added GPU continues to contribute at close to $90\%$ of
its nominal capacity. With no algorithmic coupling between GPUs, the same
driver therefore ports directly to nodes with more accelerators: descriptor
generation for ultra-large systems scales out by simply adding devices, and
a million-atom SOAP evaluation completes in under six seconds on a single
GB200 node.

\section{Conclusions}
\label{sec:conclusions}

We have presented \texttt{cuSOAP}, a GPU-accelerated generator of atom-wise
SOAP descriptors and their analytic derivatives for molecular systems. The
package rests on the closed-form working equations of
Sec.~\ref{sec:derivations} for both the Gaussian-type-orbital and the
polynomial radial bases, evaluated on the device through the fused CUDA and
\texttt{Triton} kernels of Sec.~\ref{sec:implementation}, which eliminate the
multi-gigabyte intermediates of a naive tensor-operation formulation and
reduce the derivative hot loop to a pair of fused multiply--adds per output
element. The resulting performance was quantified in Sec.~\ref{sec:results}:
for the full descriptor-plus-Jacobian workload on
$(\mathrm{H_{2}O})_{1000}$, a single Blackwell GPU outpaces the CPU-based
reference \texttt{DScribe}\cite{himanen2020,laakso2023} by up to two orders
of magnitude (i.e., $13\times$ to $227\times$ across the
$(n_{\max}, l_{\max})$ hyperparameter map, with the largest gain at the
largest basis), while reproducing its output to
within $\sim\!10^{-6}$; descriptor-only generation scales as
$t \propto n^{1.34}$, close to linear, up to a million atoms, which are
processed in $19.5$~s on one GPU and $5.7$~s on the four GPUs of a GB200
node---a parallel speedup of $3.59\times$, i.e., about $90\%$
efficiency, with no sign of saturation as devices are added.

The practical significance of these numbers lies in what they make routine.
The efficient generation of atom-wise SOAP vectors and, above all, of their
derivatives is the central bottleneck of SOAP-based molecular machine
learning: forces are required for training and running interatomic
potentials, yet their evaluation inflates the descriptor cost by a factor of
roughly $3N_{\mathrm{neigh}}$.\cite{laakso2023} At the same time, faithfully
resolving multi-species chemical environments in the condensed phase can
demand a large cutoff radius, a large radial basis $n_{\max}$, and most
critically for anisotropic coordination, a large angular band limit
$l_{\max}$.\cite{bartok2013,de2016} Precisely these physically motivated
settings are the most punishing for CPU-bound implementations, whereas the
speedup of \texttt{cuSOAP} \emph{grows} with both $n_{\max}$ and, above all,
$l_{\max}$ (Sec.~\ref{sec:results-speedup}): the angular resolution that
complex environments require, prohibitive in a CPU workflow, becomes
essentially free on the GPU. Descriptor and force evaluation for a
thousand-molecule water cluster thereby drops from half a minute to a
fraction of a second, bringing on-the-fly SOAP evaluation for
condensed-phase simulation within reach.

Adoption costs are deliberately kept low. \texttt{cuSOAP} reproduces the
constructor signature, feature ordering, and derivative index convention of
\texttt{DScribe} element for element (Sec.~\ref{sec:architecture}), so
existing pipelines can switch generators without modifying downstream code.
Its Python interface, built on \texttt{PyTorch}, accepts structures directly
as \texttt{Atoms} objects of the widely used Atomistic Simulation Environment
(\texttt{ASE}),\cite{larsen2017} through which nearly every modern
computational-chemistry and molecular-machine-learning package already
exchanges structural data; integration with existing workflows---dataset
featurization, kernel-model training, or descriptor-based analysis of
trajectories---therefore requires only a few lines of code.

Two directions will guide future development. First, the shared-memory
multiprocessing driver of Sec.~\ref{sec:multigpu} is confined to the GPUs of
a single node; because the center-partitioned workload requires no
inter-device communication, extending the same data-parallel scheme to
distributed-memory architectures via message passing is straightforward and
will allow descriptor generation for systems of tens of millions of atoms to
scale out across GPU clusters. Second, the user-facing interface will be
extended beyond Python to compiled languages such as C and Fortran, enabling
\texttt{cuSOAP} to serve as an embedded descriptor engine within established
molecular-dynamics and electronic-structure codes rather than solely as a
library for Python-based workflows.

\section*{Code availability}
Our cuSOAP code is freely available at https://github.com/TACC/cuSOAP. 

\section*{Acknowledgements}
Computational resources were provided by the Texas Advanced Computing Center at the University of Texas at Austin.

\bibliographystyle{plain}
\bibliography{references}

@article{bartok2013,
  author  = {Bart{\'o}k, Albert P. and Kondor, Risi and Cs{\'a}nyi, G{\'a}bor},
  title   = {On representing chemical environments},
  journal = {Phys. Rev. B},
  volume  = {87},
  number  = {18},
  pages   = {184115},
  year    = {2013},
  doi     = {10.1103/PhysRevB.87.184115}
}

@article{de2016,
  author  = {De, Sandip and Bart{\'o}k, Albert P. and Cs{\'a}nyi, G{\'a}bor and Ceriotti, Michele},
  title   = {Comparing molecules and solids across structural and alchemical space},
  journal = {Phys. Chem. Chem. Phys.},
  volume  = {18},
  number  = {20},
  pages   = {13754--13769},
  year    = {2016},
  doi     = {10.1039/C6CP00415F}
}

@article{laakso2023,
  author  = {Laakso, Jarno and Himanen, Lauri and Homm, Henrietta and Morooka, Eiaki V. and J{\"a}ger, Marc O. J. and Todorovi{\'c}, Milica and Rinke, Patrick},
  title   = {Updates to the {DScribe} library: New descriptors and derivatives},
  journal = {J. Chem. Phys.},
  volume  = {158},
  number  = {23},
  pages   = {234802},
  year    = {2023},
  doi     = {10.1063/5.0151031}
}

@article{himanen2020,
  author  = {Himanen, Lauri and J{\"a}ger, Marc O. J. and Morooka, Eiaki V. and Federici Canova, Filippo and Ranawat, Yashasvi S. and Gao, David Z. and Rinke, Patrick and Foster, Adam S.},
  title   = {{DScribe}: Library of descriptors for machine learning in materials science},
  journal = {Comput. Phys. Commun.},
  volume  = {247},
  pages   = {106949},
  year    = {2020},
  doi     = {10.1016/j.cpc.2019.106949}
}

@article{bigi2023,
  author  = {Bigi, Filippo and Fraux, Guillaume and Browning, Nicholas J. and Ceriotti, Michele},
  title   = {Fast evaluation of spherical harmonics with sphericart},
  journal = {J. Chem. Phys.},
  volume  = {159},
  number  = {6},
  pages   = {064802},
  year    = {2023},
  doi     = {10.1063/5.0156307}
}

@article{behler2007,
  author  = {Behler, J{\"o}rg and Parrinello, Michele},
  title   = {Generalized neural-network representation of high-dimensional potential-energy surfaces},
  journal = {Phys. Rev. Lett.},
  volume  = {98},
  number  = {14},
  pages   = {146401},
  year    = {2007},
  doi     = {10.1103/PhysRevLett.98.146401}
}

@article{bartok2010,
  author  = {Bart{\'o}k, Albert P. and Payne, Mike C. and Kondor, Risi and Cs{\'a}nyi, G{\'a}bor},
  title   = {Gaussian approximation potentials: The accuracy of quantum mechanics, without the electrons},
  journal = {Phys. Rev. Lett.},
  volume  = {104},
  number  = {13},
  pages   = {136403},
  year    = {2010},
  doi     = {10.1103/PhysRevLett.104.136403}
}

@article{deringer2021,
  author  = {Deringer, Volker L. and Bart{\'o}k, Albert P. and Bernstein, Noam and Wilkins, David M. and Ceriotti, Michele and Cs{\'a}nyi, G{\'a}bor},
  title   = {Gaussian process regression for materials and molecules},
  journal = {Chem. Rev.},
  volume  = {121},
  number  = {16},
  pages   = {10073--10141},
  year    = {2021},
  doi     = {10.1021/acs.chemrev.1c00022}
}

@article{larsen2017,
  author  = {Larsen, Ask Hjorth and Mortensen, Jens J{\o}rgen and Blomqvist, Jakob and Castelli, Ivano E. and Christensen, Rune and Du{\l}ak, Marcin and Friis, Jesper and Groves, Michael N. and Hammer, Bj{\o}rk and Hargus, Cory and Hermes, Eric D. and Jennings, Paul C. and Jensen, Peter Bjerre and Kermode, James and Kitchin, John R. and Kolsbjerg, Esben Leonhard and Kubal, Joseph and Kaasbjerg, Kristen and Lysgaard, Steen and Maronsson, J{\'o}n Bergmann and Maxson, Tristan and Olsen, Thomas and Pastewka, Lars and Peterson, Andrew and Rostgaard, Carsten and Schi{\o}tz, Jakob and Sch{\"u}tt, Ole and Strange, Mikkel and Thygesen, Kristian S. and Vegge, Tejs and Vilhelmsen, Lasse and Walter, Michael and Zeng, Zhenhua and Jacobsen, Karsten W.},
  title   = {The atomic simulation environment---a {Python} library for working with atoms},
  journal = {J. Phys.: Condens. Matter},
  volume  = {29},
  number  = {27},
  pages   = {273002},
  year    = {2017},
  doi     = {10.1088/1361-648X/aa680e}
}

\end{document}